# Carrier free long-range magnetism in Mo doped one quintuple layer $Bi_2Te_3$ and $Sb_2Te_3$

Xiaodong Zhang[1,2] and Junyi Zhu[2*]

[1] Beijing Computational Science Research Center, Beijing 100193, China

[2] Department of Physics, Chinese University of Hong Kong, Hong Kong SAR

Abstract: One of the keys to the realization of Quantum Anomalous Hall effect (QAHE) is long range ferromagnetism, which is only experimentally realized in Cr or V doped $(Bi,Sb)_2Te_3$ system. Both elements are $3d$ transition metals and $4d$ transition metals are found to be ineffective to produce long range ferromagnetism in $Bi_2Se_3$. Still, whether long range ferromagnetism can be realized by magnetic doping of $4d$ elements is an open question. Based on density functional theory calculations, we predict that long range ferromagnetism can be realized in Mo doped $Bi_2Te_3$ and $Sb_2Te_3$, which are semiconducting. The coupling strength is comparable with that of Cr doped $Bi_2Te_3$ and $Sb_2Te_3$. Therefore, Mo doped $Bi_2Te_3$ and $Sb_2Te_3$ or their alloys can be new systems to realize diluted magnetic semiconductors and QAHE.

## 1. Introduction

Long range magnetic interaction is the key to the realization of magnetic semiconductors and many exotic physics. Recently, the most important success in carrier free long range ferromagnetism is the finding of Quantum Anomalous Hall effect (QAHE) in Cr or V doped $(Sb,Bi)_2Te_3$ [1,2]. Yet no any other transition metal dopants have been reported to realize long range magnetism.

In addition to direct doping of transition metals, proximity effects have been found to induce long range ferromagnetism [3–5]. Still, the realization of proximity effects requires thin films with very strong magnetism and defect free interfaces between the topological insulator layers and the magnetic layers. Therefore, direct doping of transition metal is still the preferred approach in many studies [6–13].

There are three major obstacles in the magnetic doping: (1) limited choices of dopants, (2)

* jyzhu@phy.cuhk.edu.hk

relatively weak ferromagnetic coupling, and (3) clustering of magnetic dopant.

Isovalent dopants are the top candidates because insulated bulk with topologically non-trivial surface states is needed to realize QAHE. Usually, transition metal elements with a valence of three are incorporated to replace a cation in $(Sb_{1-x}Bi_x)_2Te_3$. Still, limited dopants have been found to have three valence electrons in the electronic environment of the tellurides. Currently, the candidates include Cr [1,14], Fe [14] and Gd [15]. To overcome the isovalent doping difficulty, codoping has been proposed, e.g. V and I codoped $Sb_2Te_3$ [18]. Still, codopants may introduce detrimental defects [19] and be corrosive to growth chambers [17]. In addition to $3d$ transition metals, $4d$ transition metal elements can be potential candidates. However, early theoretical calculations predicted that $4d$ metals may introduce a high $d$ level above the Fermi level of $Bi_2Se_3$, thus introduce detrimental gap states that destroy the insulating nature of the bulk material. Whether $4d$ transition metals dopants in the tellurides are isovalent is largely unknown.

Ideally, strong magnetic interaction is preferred for QAHE to achieve high transition temperature. In reality, weak ferromagnetic interaction is often the case. Vergniory et al. have systematically studied the $3d$ transition metal doped $Bi_2Se_3$ family and found that only Ti, V, Cr, and Mn prefer ferromagnetic state [18]. Among them, Mn doped $Sb_2Te_3$ have very weak magnetic interaction [20]. Recently, systematic experimental study also suggested very weak or no ferromagnetic order in the system [19]. The Fe and Co prefer antiferromagnetic states, verified by experiments [20,21]. The small ferromagnetic coupling strength or even anti-ferromagnetism limit the choice of magnetic dopants to realize long range ferromagnetism.

The segregation of magnetic dopant, i.e. clustering, can hinder the formation of long-range magnetic order. The disappearance of long-range magnetic order due to clustering was firstly experimentally demonstrated in Cr doped $Bi_2Se_3$ [22]. Later, various experiments showed existence of long-range magnetic order in Te-based systems [1,2,23–25]. Our early theoretical investigations explained the existence of long range magnetic interaction that stabilizes the long range separation of magnetic dopants in Cr in $Bi_2Te_3$ and $Sb_2Te_3$ systems [26]. Clustering of magnetic dopants is also found in Mn- or Co-doped $Bi_2Te_3$ [21,27]. Therefore, Cr is so far the only intrinsic magnetic dopant with long range magnetic interaction that is strong enough to suppress clustering [26]. So, suppression of clustering is the most important prerequisite to realize QAHE.

In this paper, based on our density functional calculations, we propose to dope Mo in $Bi_2Te_3$ and $Sb_2Te_3$ to solve the aforementioned problems. Mo is a promising candidate because it belongs to XIB group, the same as Cr. Although the gap states due to the $d$ states of Mo were reported in Mo doped $Bi_2Se_3$ [28], the relative high $p$ states of Te comparing to that of Se suggest that Mo doped tellurides can be free of detrimental gap states. Therefore, we calculated the electronic properties, magnetic interactions, and formation energy of Mo doped $Bi_2Te_3$ and $Sb_2Te_3$ and found that carrier free long-range magnetic order is indeed stable in both systems. Moreover, the ferromagnetic coupling strength is larger than that of Cr doping case.

## 2. Computational method and details

All calculations were performed using projected augmented wave (PAW) [29,30] potentials with Perdew-Burke-Ernzerhof (PBE) [31] generalized gradient approximation (GGA) as implemented in Vienna *ab initio* simulation package (VASP) [32]. The cutoff energy for plane-wave expansion was set to 400eV for both $Bi_2Te_3$ and $Sb_2Te_3$. Gamma centered 4×4×1 $k$ mesh was used to sample the Brillouin zone for 4×4 supercell of $Bi_2Te_3$ and $Sb_2Te_3$, as shown in Fig. 1, with all the neighboring sites marked. The number 0 represents the position of first Mo or Cr atom. The numbers 1 to 8 represent the 1st to 8th nearest neighbor (NN) placements of the second Mo or Cr atom. All atoms in every supercell are fully relaxed until the residual force is less than 0.01 eV/Å. For PBE+U calculations, parameters of $U$=3.6eV, $J$=0.6eV are taken and part of the results are compared with those of HSE06 calculation. Convergence tests about $k$-points, cell sizes, vacuum size, magnetism, and energy cutoffs have been performed. The formation energy of the dopants is defined as:

$$\Delta H_f(X) = E_{tot}(M_2Te_3:X) - E_{tot}(host) - \Sigma_i n_i \mu_i \qquad (1),$$

where $E_{tot}$ ($M_2Te_3$:X) is the total energy of a supercell of $M_2Te_3$ (M=Bi or Sb) with X (X=Cr or Mo) dopants; $E_{tot}$ (host) is the total energy of the supercell without impurities; $n_i$ is the number of certain atoms added to ($n_i > 0$) or removed from ($n_i < 0$) the supercell; and $\mu_i$ is the corresponding chemical potential. More details about the formation energy calculations are shown in Supplementary materials. In the main text, we take the relative formation energy in reference to the lowest energy among all configurations.

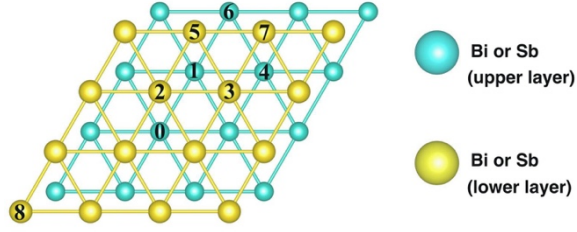

**Figure 1**: Schematic illustration of one quintuple layer X$_2$Te$_3$ (Bi or Sb) supercell (4×4) with blue and yellow atoms being the X in the upper and lower layers, respectively. The Te atoms are removed for clarity.

## 3. Results and discussions

### 3.1 Formation energy of single magnetic dopant

In order to determine the feasibility of doping transition metal, X, in M$_2$Te$_3$, we calculated the formation energy of single X atom substituting M. The formation energy is defined as following:

$$\Delta H_f(X_M) = E_{tot}(doped) - E_{tot}(host) + \mu_M - \mu_X \quad (2)$$

where $E_{tot}(doped)$ is the energy of single doped telluride, $E_{tot}(host)$ is the energy of undoped systems, $\mu_M$ and $\mu_X$ are the chemical potential of M and X, respectively. The chemical potential $\mu_X$, $\mu_{Te}$ and $\mu_M$ are variables, however, they have to fulfill the following conditions to keep the host materials stable and avoid formations of secondary phases of XTe$_2$:

$$E_{tot}(M_2Te_3) = 2\mu_M + 3\mu_{Te} = 2\mu_M^{solid} + 3\mu_{Te}^{solid} + \Delta H_f(M_2Te_3) \quad (3)$$

$$\mu_X + 2\mu_{Te} \leq E_f(XTe_2) \quad (4).$$

where the $\mu_M^{solid}$ and $\mu_X^{solid}$ are the total energy of ground state solid M and X, respectively. $\Delta H_f(M_2Te_3)$ is the formation energy of M$_2$Te$_3$.

Also, to compare with the doping of Cr that was experimentally achieved, the formation energy of Cr was also calculated. According to calculated results (Fig. 2), the formation energy of Mo under Te-rich condition is smaller than that under Te-poor condition. This result is consistent with previous experimental results and theoretical predictions that magnetic doping is energetically preferred under anion rich conditions [1,2,33]. Besides, the incorporation of magnetic dopants is easier in Sb$_2$Te$_3$ than that in Bi$_2$Te$_3$. These trends are also consistent with previous studies about 3$d$ transition metals doping. For Mo doped Bi$_2$Te$_3$, although the formation energy of Mo doping case is larger than that of Cr doping case, it is still possible to incorporate Mo, as the formation energy of Mo doped Bi$_2$Te$_3$ is negative under Te-rich condition [Fig. 2(a)]. The large negative formation energy of Mo or Cr dopants suggests that both dopants can be spontaneously incorporated [Fig. 2(b)].

Earlier studies suggest that the formation energy can be influenced by spin orbital coupling (SOC) in 3$d$ transition metal doped Bi$_2$Se$_3$. So, it is essential to study such effects for both Bi$_2$Te$_3$ and Sb$_2$Te$_3$ systems. We found that the SOC largely reduces the formation energy of Bi$_2$Te$_3$ case (about 43% and 46% for Cr and Mo, respectively) and slightly increases the formation energy of $Sb_2Te_3$ case (no more than 8%). The formation energy of Mo in $Bi_2Te_3$ is largely reduced, comparing with that in $Sb_2Te_3$, because the relatively large SOC effect on Bi may push down the *p* state of Bi and enhance the effective *p-d* coupling between Bi and Mo. Similar trend is also found in other 3*d* transition metal doped systems [33]. Considering the negative formation energy for Te-rich condition, we still expect that the Mo can be doped in both $Bi_2Te_3$ and $Sb_2Te_3$ systems.

For Cr doping case, it should be noted that the formation energy of our calculations in 1QL model is smaller than that of previous studies in bulk systems [34]. This difference is expected as structural relaxation around magnetic dopants in 1QL is much easier than that in bulk systems because 1QL system has more free room to relax the stress induced by the dopants. Therefore, it is expected that the doping of magnetic dopants should be performed in monolayer or few layer thin films.

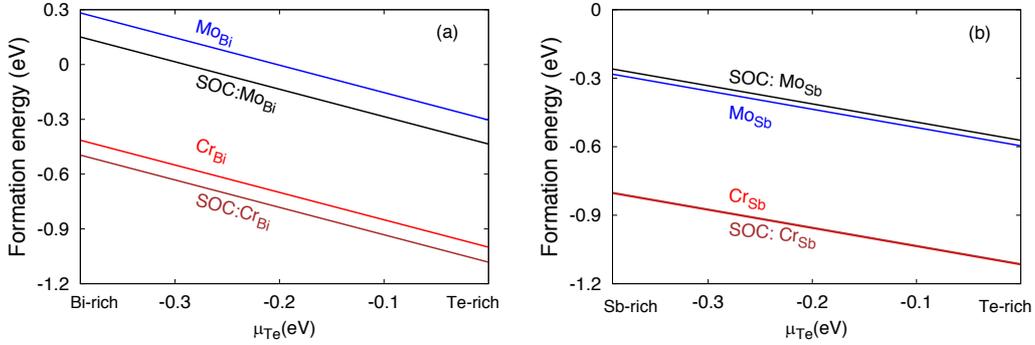

Fig. 2 Formation energy of single Cr/Mo doped $Bi_2Te_3$ and $Sb_2Te_3$ as chemical potential of Te.

**3.2 Electronic structure of single magnetic dopant**

To reveal the electronic structure of Mo doped $Bi_2Te_3$ and $Sb_2Te_3$, we calculated the density of states (DOS) and magnetic moments by PBE+U and HSE06 methods. Considering the similarity of $Bi_2Te_3$ and $Sb_2Te_3$, here we only show the DOS of Mo doped $Bi_2Te_3$ as an example (Fig. 3).

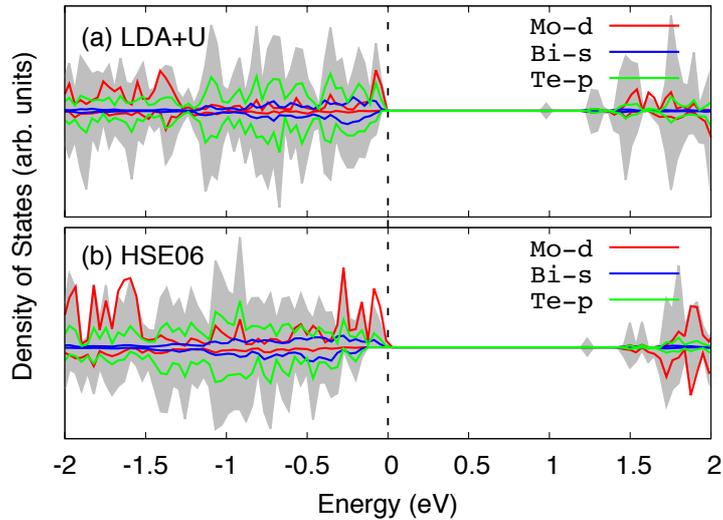

**Fig. 3** Density of states of Mo doped $Bi_2Te_3$. (a) PBE+U results (b) HSE06 results. The gray

background is the total density of states, and the blue line denotes *d* orbitals of Mo.

Contradicting to Mo doped $Bi_2Se_3$ [28], the system remains insulating upon Mo doing and no impurity state is found within the gap for both PBE+U and HSE06 calculations [Fig.3]. This difference is not unexpected as the energy levels of Te-*p* orbitals are higher than that of Se-*p* orbitals. Such trend can be found in other transition metal doped $Bi_2Se_3$ family [35]. So, the substitution of Mo for Bi and Sb should be isovalent doping.

The Mo-*d* orbitals below the Fermi level are hybridized with intrinsic *s-p* orbitals of host systems. Also, states derived from *s-p* orbitals are spin-polarized, which may provide the essential bridge to mediate the long range magnetic interaction [26]. Specifically, similar to the Cr doping case, the magnetized *s* orbital of Bi may play an important role in mediating the magnetic interaction between two Mo dopants, as illustrated by the pDOS of Bi state near the valence band maximum (VBM), and the magnetic mechanism associating with this scenario will be presented below.

Also, both methods produce the same magnetic moments (3.0$\mu$B), again indicating PBE+U is accurate enough to describe the magnetic properties of the doped telluride. Considering the approximately $O_h$ local symmetry, *d* orbitals are split into filled majority $t_{2g}$ orbitals and empty majority $e_g$ orbitals. Due to the large crystal splitting, both minority $t_{2g}$ orbitals and $e_g$ orbitals are empty. Therefore, the magnetic moments should be contributed by the occupied $t_{2g}$ orbitals.

**3.3 Relative formation energy**

Further, we calculate the relative formation energy of two dopants in $Bi_2Te_3$ and $Sb_2Te_3$. The relative formation is defined by the formation energy difference between every configuration and the FM state of 7th-NN and the results are shown in Fig. 4.

The energy difference between short-range configurations (1st-NN and 2nd-NN) and long-range configurations (e.g. 7th-NN) $\Delta E_{1st-7th}$ is an important factor determine the formation of magnetic clusters. Comparing with Cr doping case ($\Delta E_{1st-7th} = 5.4 meV$), the energy difference is 48.7meV and 53.8meV for Mo doped $Bi_2Te_3$ and $Sb_2Te_3$, respectively. The suppression of short-range configurations can largely improve the homogeneity. It is found that the stability of QAH state and Curie temperature is not only dependent on magnetic interaction, but also dependent on homogeneity of magnetic dopant [36]. Thus, we expect that the observed critical temperature of QAHE in Mo doping case should be larger than that of Cr doping case because of the improved homogeneity and comparable magnetic interaction. In addition, it should be noted that the formation of isolated magnetic clusters can be largely suppressed in thin film comparing to bulk system [26]. Therefore, we expect that the homogeneity of Mo doping can be improved in few-layer thin films.

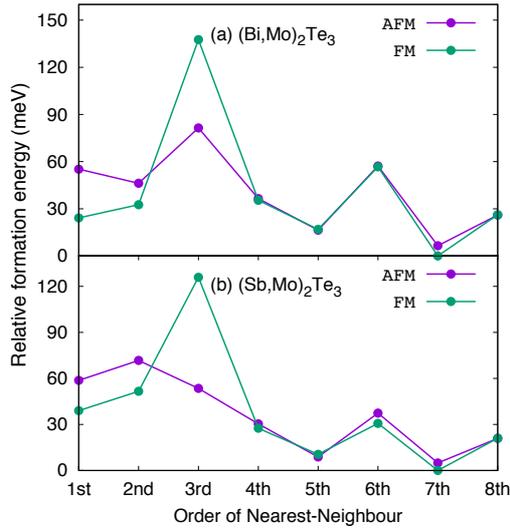

**Fig. 4** Relative formation energy of Mo doped (a) $Bi_2Te_3$ and (b) $Sb_2Te_3$.

### 3.4 Magnetic interaction

Here, the ferromagnetic interaction is defined by energy difference between FM state and AFM state. Large enough long-range ferromagnetic interaction is essential to realize long range ferromagnetism. So, we firstly calculate the ferromagnetic interaction of two Mo atoms in various sampling configurations and the results are shown in Fig. 5. It is found that the ferromagnetic interaction is robust in long range configurations, e.g. 6th-NN and 7th-NN. For 7th-NN, the ferromagnetic coupling strength is about -7meV and -6meV for Mo doped $Bi_2Te_3$ and Mo doped $Sb_2Te_3$, respectively.

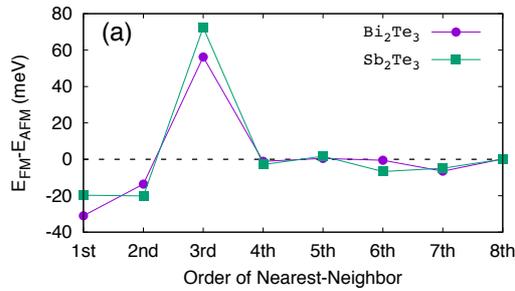

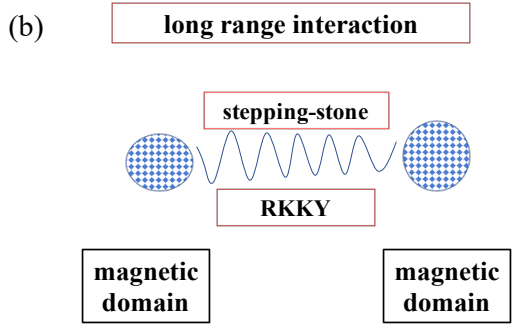

Fig. 5. (a) The magnetic interaction in various nearest-neighbor configurations and (b) schematic illustration of long-range magnetic interaction between magnetic domains.

Usually, the magnetic clusters are inevitable in diluted magnetic semiconductors. With magnetic clusters formed, the long-range magnetic interaction can be mediated by carriers *via* RKKY or Zener's model. However, there is no carrier in the bulk of magnetic doped topological insulators. Thus, the carrier free long-range magnetic interaction (e.g. stepping-stone mechanism) may play a critical role to form overall ferromagnetism, as shown in Fig.5(b).

The magnetic properties are sensitive to correlation effects of transition metals and spin-orbital coupling (SOC), so, we further calculated the ferromagnetic coupling strength of 7th-NN configuration by PBE+U and HSE06 method with and without SOC. We found that the result of PBE+U is consistent with HSE06 result (Table 1). The similar ferromagnetic coupling strength again endorse the validity of PBE+U method. Also, the ferromagnetic coupling strength is comparable to the value of Cr doping case [26], thus, we expect that long-range ferromagnetism is possible to realize in Mo doped tellurides.

When SOC effect is included, the ferromagnetic state still remains although the ferromagnetic coupling strength is somewhat smaller. This is not surprising because the magnetized *s-p* states near the VBM still can mediate the magnetic interactions although the SOC changes the relative positions of the Te-*p* orbitals. This trend is similar to our previous results of Cr doped $Bi_2Se_3$ family [26].

Table 1 The ferromagnetic coupling strength of 7th-NN in Cr or Mo doped $Bi_2Te_3$

|  | **FM-AFM (meV)** |
|---|---|
| PBE+U for Cr doping [26] | -6.2 |
| PBE+U | -6.5 |
| HSE06 | -6.7 |
| HSE06+SOC | -3.2 |

**3.5 Magnetic coupling mechanism**

Spin density can provide hints for clarifying the magnetic mechanism. According to the spin density of Mo doped $Bi_2Te_3$ (Fig. 6), we found that the spin of Mo-*d* is antiparallel to that Te-*p*. Further, the hybridized state formed by Bi-*sp* and Te-*p* at the stepping stone site [26] is magnetized, and this hybridized state couples with the second Mo dopant. The *s-p* network provides a medium for Mo-Mo interaction. This is the first system other than Cr doped ones [26] that verified the existence of ferromagnetism mediated by stepping stone mechanism.

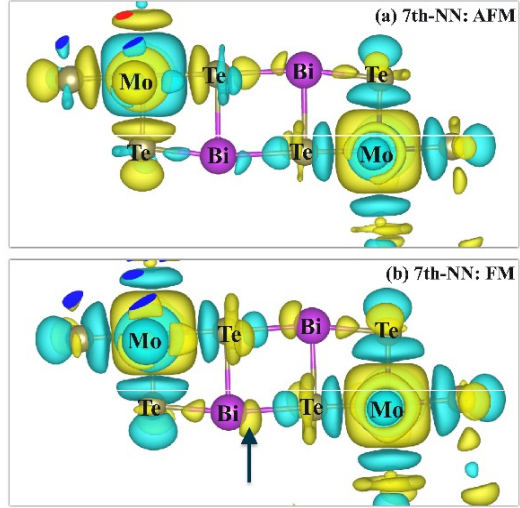

Fig. 6. Spin density of 7th-NN configuration in different magnetic states, (a)AFM state and (b) FM state. The *sp* antibonding state is marked by the arrow.

In magnetic doped TI, various mechanism can mediate the long range ferromagnetism, such as RKKY [37], van vleck mechanism [38], double or superexchange mechanism [18], *p*-network mediated mechanism [39], and stepping stone mechanism [26]. In Mo doped $Bi_2Te_3$ and $Sb_2Te_3$, no impurity state was found within the band gap, therefore, the possibility of RKKY is excluded. According to van vleck mechanism, SOC is essential to realize long range magnetic coupling [14]. However, our calculations found that the magnetic coupling strength remains without SOC. Double exchange mechanism involve difference valence of magnetic ions, which is absence in our system. For superexchange or super-superexchange [40], the interaction is usually short ranged. However, we found that magnetic interaction is still robust in 7th-NN configurations (about 10Å). Above analysis indicates that the stepping stone mechanism is likely to play the dominant role.

In addition, it was suggested that the magnetic interaction can be mediated by the topological surface state *via* RKKY like mechanism [41]. However, our present calculation is based on 1QL model, in which the band gap is clean without topological surface states because of the strong coupling between the top and bottom surface states of the thin film [42]. Therefore, the role of topological surface states for the long-range magnetic mechanism is absent here. Further studies based on 5QL model or thicker thin film can be interesting and may clarify the role of topological surface state on magnetism, which is out of the scope of this paper.

### 4. Summary

Based on density functional theory, we calculated the electronic structure, formation energy and magnetic interaction of Mo doped $Bi_2Te_3$ and $Sb_2Te_3$. The Mo doping induces ferromagnetic ground state with low formation energy. Moreover, the clustering is highly suppressed in Mo doping case. These results suggest that the Mo doped $Bi_2Te_3$ and $Sb_2Te_3$ can provide a new platform to realize diluted magnetic semiconductors and the QAHE related exotic quantum phenomenon.

**Acknowledgement**

X.Z and J.Y.Z are grateful for the financial support of Chinese University of Hong Kong (CUHK) (Grant No.4053084), University Grants Committee of Hong Kong (Grant No. 24300814), and start-up funding of CUHK. X.Z acknowledges financial support from NSFC by grant No.11734002, NSAF by grant No.U1530401 and computational resources from the Beijing Computational Science Research Center.